\documentclass[fleqn,twoside]{article}
\usepackage{espcrc2}

\usepackage{graphicx}
\usepackage{epsfig}

\newcommand{\AmS}{{\protect\the\textfont2
  A\kern-.1667em\lower.5ex\hbox{M}\kern-.125emS}}

\title{ A phase transition due to thick vortices in
       SU(2) lattice gauge theory}

\author{Srinath Cheluvaraja\address{Department of Physics and Astronomy,
Louisiana State University,
        \\ 
        Baton Rouge, USA }%
        \thanks{ This research was supported in part by United States
Department of Energy grant DE-FG 05-91 ER 40617.
                }}
\begin{document}
\def \beq {\begin{equation}}
\def \eeq {\end{equation}}

\begin{abstract}
$SU(2)$ lattice gauge theory is studied after eliminating thin monopoles
and the smallest thick
monopoles. Kinematically this constraint
allows thick vortex loops which produce long range $Z(2)$
fluctuations.
The thick
vortex loops are identified in a three dimensional simulation. A
condensate  of thick vortices persists even after the
thin vortices have all disappeared. They decouple
at a slightly lower temperature (higher $\beta$) than
the thin vortices and drive a $Z(2)$ like phase transition.

\vspace{1pc}
\end{abstract}

\maketitle
\section{Introduction}
The role of thick vortices \cite{mack79,tombo81} as a mechanism for
confinement in $SU(2)$ LGT has been extensively discussed in the last
few years \cite{latts}.
Thick
vortices are analogous to the domain walls in ferromagnets with a
continuous symmetry--they are like thick Peierls contours.
Thick vortices should be distinguished from thin vortices.
Thin vortices
have infinite action in the continuum limit. For this reason they
are unlikely to play a major role in the continuum limit.
On the other hand
thick vortices can reduce their free energy by arbitrarily increasing their
cross section \cite{kt00}.
In addition to thin and thick vortices
$SU(2)$ gauge theory also has thin and thick monopoles.
A thin vortex can
end in a thin $Z(2)$ monopole which (in three space-time dimensions) is defined on an elementary 3 dimensional 
cube.
The thin $Z(2)$ monopole density 
on a cube $c$ is
given by
\beq
\rho_{1}(c)=(1/2)(1-sign(\prod_{p \in \partial c} trU(p)))
\quad .
\eeq
Analogously, a thick vortex can end in a thick $Z(2)$ monopole.
Its density (in three space-time dimensions) is given by
\beq
\rho_{d}(c_{d})=(1/2)(1-sign(\prod_{d \in \partial c_{d}} trU(d)))
\quad ,
\eeq
where the product is taken over all $dXd$ loops bordering a cube 
$c_{d}$ of
side $d$.
The subscript $d$ indicates that the density can be defined on any
3 dimensional cube of side $d$.
In three space-time dimensions the monopoles are point like objects or
they have a finite size.
Both monopoles are $SO(3)$ invariant as they only
depend on the $SU(2)/Z(2)$ coset of the link variables.
A whole
hierarchy of such monopoles can be defined with different thicknesses.
Unlike the thin monopoles, which are suppressed by the action, the thick
monopoles cost lesser energy because their energy is spread over a finite
region.
It was proposed  in \cite{mack80} that at lower and lower temperatures
vortices and monopoles of thicker and thicker cross-section are
important.
These ideas were encapsulated
in the effective $Z(2)$ theory of confinement 
\cite{mack80} where it was proposed that
the long distance properties of
$SU(2)$ LGT can be described by a
$Z(2)$ theory with an effective coupling $\beta(d)$ ( $\beta(d) \rightarrow
\infty \  as \  d \rightarrow \infty$). Thick vortices are the vortices
in the effective $Z(2)$ theory.

The presence of thick monopoles prevents the thick vortices from forming
closed loops and makes it difficult to study their effects separately.
If the thick monopoles are removed it might be possible
to see if thick vortices are
present at all and then study their effects. With this aim in mind we
study the $SU(2)$ theory after eliminating the thin and
thick monopoles. This model can be regarded as a generalisation of
the Mack-Petkova model which eliminates only the thin monopoles.
Just as in the Mack-Petkova model the elimination of the thin and thick
monopoles changes only the short distance properties of the gauge theory.
\includegraphics[width=15pc,height=15pc]{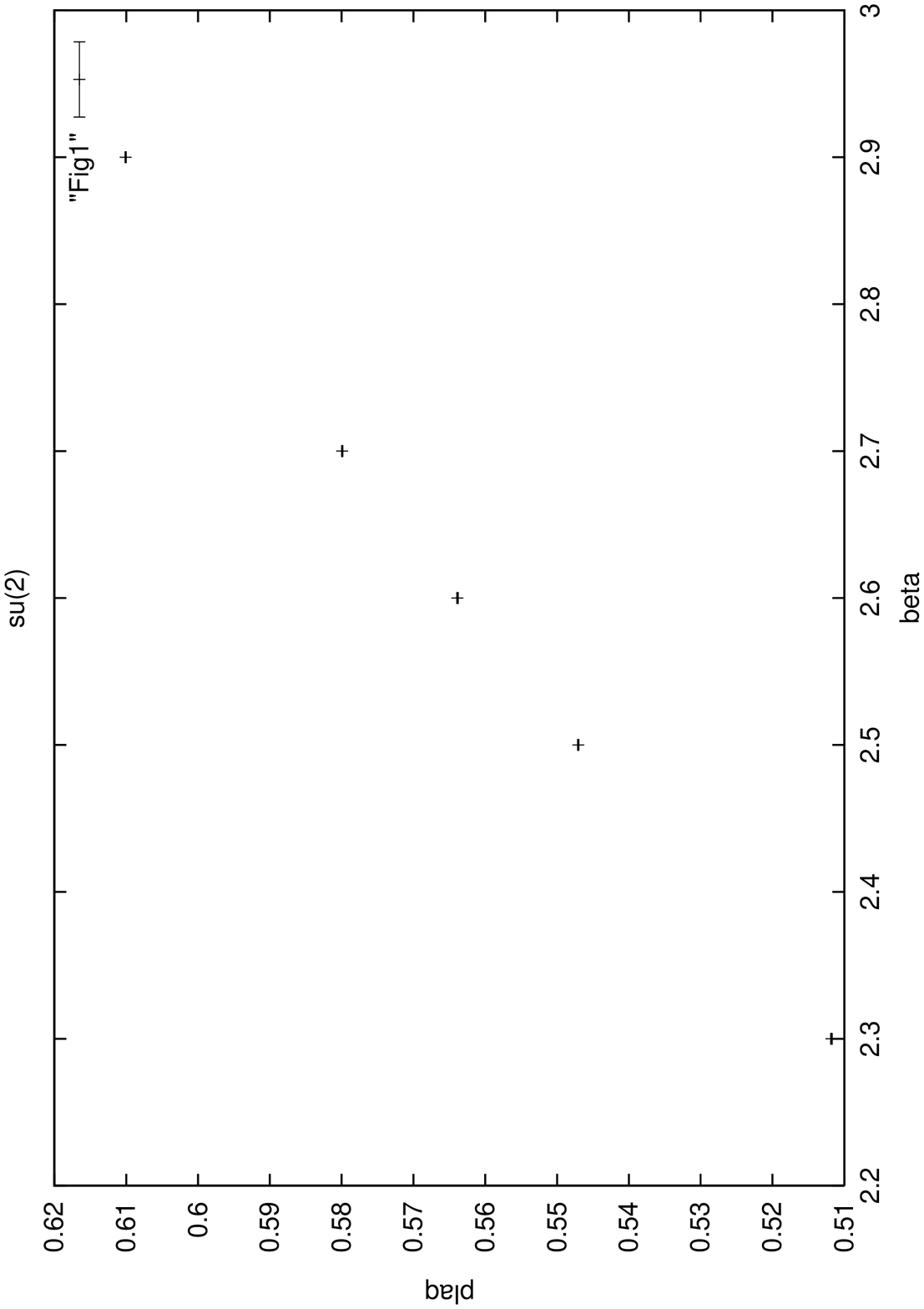}

\noindent
Fig.1 The $2X2$ Wilson loop in the $SU(2)$ theory.
\section{The model}
Thin and thick monopoles can be removed 
by introducing two (large) chemical potential
terms for the thin and thick monopole densities in the
Wilson action.
The action considered is
\beq
S=\frac{\beta}{2}\sum_{p}tr\ U(p) -\lambda_{1}\sum_{c_{1}}\rho_{1}(c_{1})
-\lambda_{2}\sum_{c_{2}}\rho_{2}(c_{2})
\quad .
\eeq
$\lambda_{1},\lambda_{2}$ are chosen to be large large so that the monopole
densities are very small ( in practice $\lambda_{1}=\lambda_{2}\approx 20$
resulted in identically zero densities for almost all thermalised 
configurations on
$12^3$ lattices).

The simulation of this model presented its own share of difficulties.
The environment of a single link is complicated because each
link touches four thin monopoles and eight thick monopoles. Different
strategies were attempted to simulate this model. Metropolis updating and
a combination of heat bath and metropolis were both tried but
metropolis updating was found to be more efficient provided the
table of $SU(2)$ elements is tuned  regularly to get a reasonable acceptance.
One observes
long metastabilities while thermalising the lattice and the
simulation has to be kept running for unusually long times unless a
good starting configuration is chosen. It was found that in the phase
where thick vortices are in abundance an ordered start (which has zero
density of monopoles and vortices)
took a very
long time to reach the equilibrium distribution. Similarly, in the phase
where thick vortices are absent a random start ( which has a large number
of thick vortices) took a long time to reach the equilibrium
distribution. 
\includegraphics[width=15pc,height=15pc]{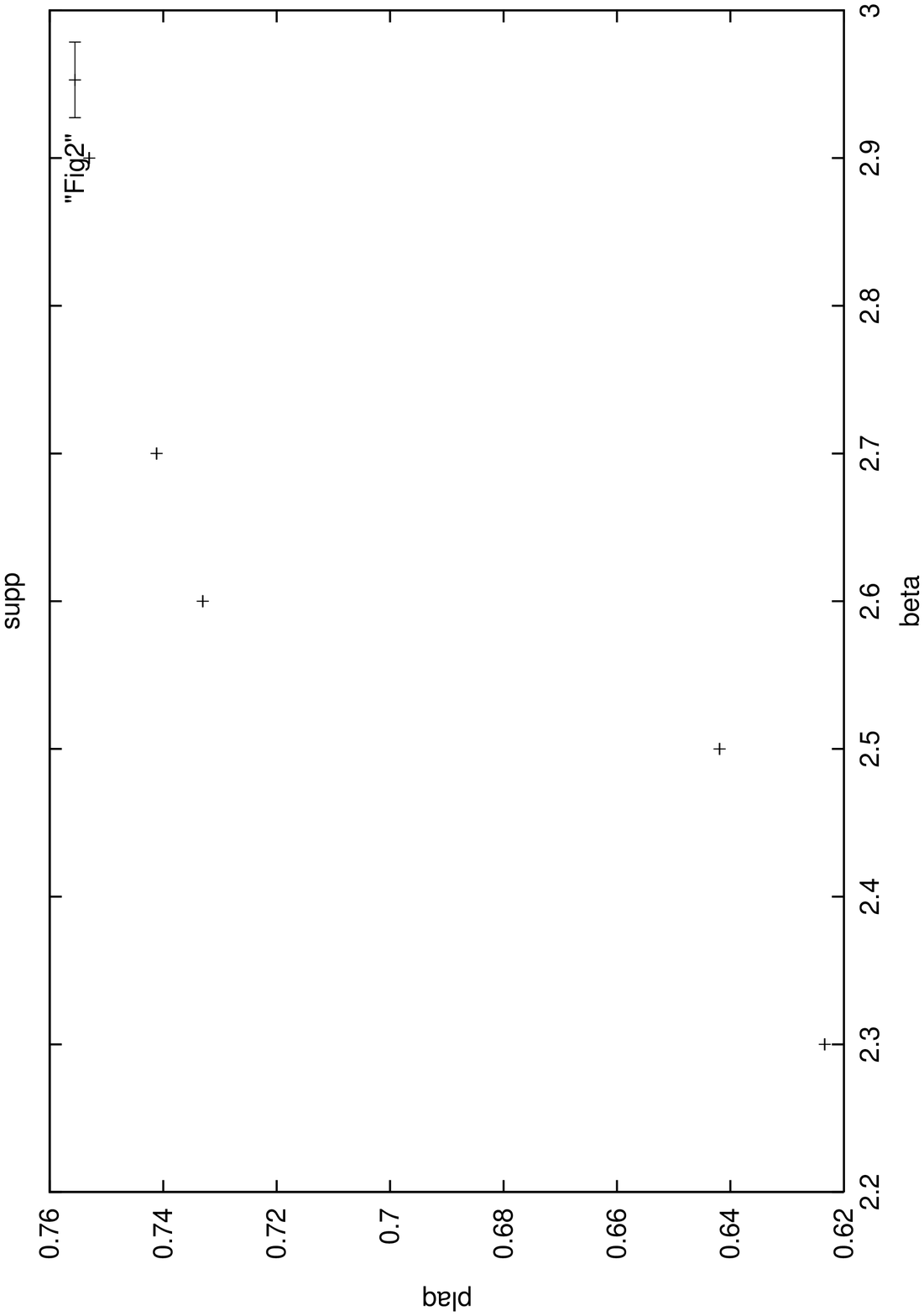}

\noindent
Fig2. The plaquette in the monopole suppressed model. 

\noindent
These metastabilities donot arise in the pure $SU(2)$
theory or the Mack-Petkova model and they appear to be linked to the
thick vortices present in this model.
Several approaches were tried to deal with this metastability like
starting from configurations "closer" to the equilibrium configuration
but the problem always remained. Nevertheless, with increased computational
effort consistent results can be and were obtained.
The simulation was performed in
three dimensions mainly for reasons of computational speed.
On the lattices used ($12^3$) it was found that simulating this model
is more time consuming than simulating the four dimensional pure 
$SU(2)$ model.

\section{ Numerical results}
It is well known that 3 dimensional $SU(2)$ LGT does not exhibit the
crossover found in the four-dimensional theory but instead has a very
smooth behaviour for the plaquette and other quantities. 
In Fig 1 we show this smooth behaviour in the $2X2$ Wilson loop.
Eliminating
thin monopoles leads to a slight jump in the plaquette due to the
presence of thin vortices. Eliminating the thick monopoles leads to
another jump at a lower temperature. Our study has focused on the
properties of the system around this point. Fig 2 and Fig 3 show the
sharp jump in the plaquette and the $2X2$ Wilson loop respectively.
In the vicinity of this jump very long relaxation times are observed
before a starting configuration reaches equilibrium.
The sign of the $2X2$ Wilson loop changes abruptly across
this transition. It should be emphasized that the thin vortices have
all disappeared at this point 
 (the sign of the plaquette is always a positive quantity here) and
the fluctuations present are over longer distances.

Since $Z(2)$ fluctuations at longer length scales 
are observed it is natural
to ask if these fluctuations are caused by thick vortices. 
Indeed they are! Thick vortex loops can be identified as closed loops
which are composed of a co-closed set of $2X2$ plaquettes with a
negative sign.
A measurement
of the thick vortices on either side of the transition shows that they
form very long loops in one phase and smaller loops in the other
phase. Hence, the thick vortices are producing the $Z(2)$ fluctuations
at longer distances, in line with the ideas of 
the effective $Z(2)$ theory of
confinement.
The location of the phase transition was determined to
be around $\beta \approx 2.5$.
\includegraphics[width=15pc,height=15pc]{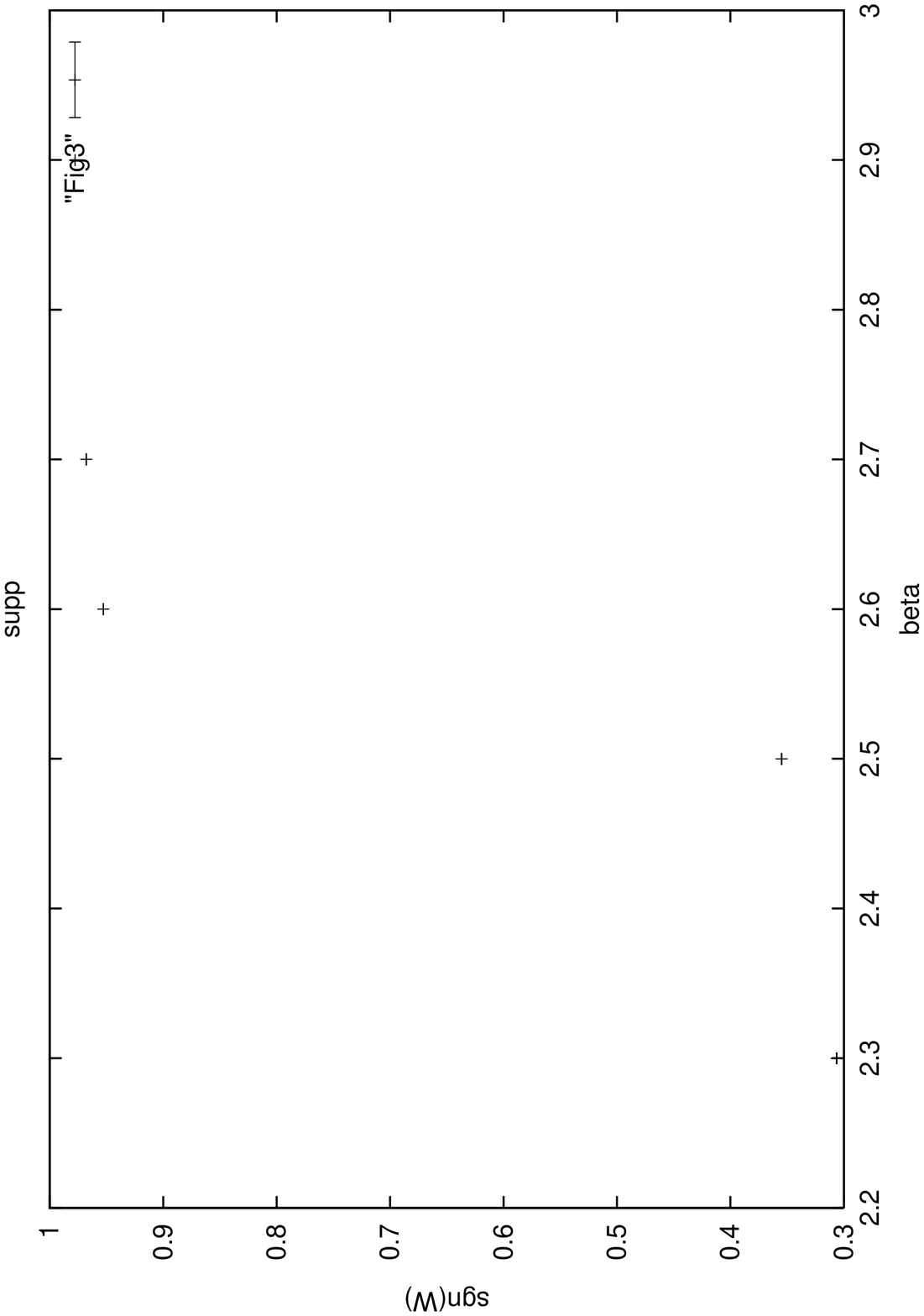}

\noindent
Fig3. The sign of the $2X2$ Wilson loop in the monopole suppressed model.

\section{Conclusions}
Thick vortices can be directly observed once the thick
monopoles are eliminated. The condensate of thick vortices persists
at higher values of $\beta$ after the thin vortices have all disappeared.
They are responsible for the long range $Z(2)$ fluctuations which are
a crucial ingredient in the effective $Z(2)$ theory of confinement.
As $\beta$ increases
the lattice spacing becomes 
smaller and vortices with a non-zero
physical thickness can still be present.
In the lattice model
there will be a hierarchy of effective $Z(2)$ theories operating at
larger and larger $\beta$ \cite{mack80}; the model studied here is just
the first member of the hierarchy.
Eliminating monopoles of greater thickness 
will in principle unravel
the entire hierarchy of $Z(2)$ like theories operating at larger
lattice spacings . It is a challenge to show that in the continuum limit
we are left with $Z(2)$ fluctuations at physical
distances of the order of a
fermi as predicted in \cite{kt00}.

\end{document}